\documentclass[10pt,conference, dvipsnames]{IEEEtran}
\IEEEoverridecommandlockouts
\usepackage{cite}
\usepackage{amsmath,amssymb,amsfonts}
\usepackage{float}
\usepackage{mathtools}
\usepackage{algorithmic}
\algsetup{linenosize=\small}
\usepackage{graphicx} \graphicspath{ {Figures/} }
\usepackage{color}
\usepackage{xcolor}
\usepackage{textcomp}
\def\BibTeX{{\rm B\kern-.05em{\sc i\kern-.025em b}\kern-.08em
    T\kern-.1667em\lower.7ex\hbox{E}\kern-.125emX}}
\usepackage[ruled,linesnumbered,vlined]{algorithm2e}
\usepackage{multirow}
\usepackage{comment}
\usepackage{booktabs} 
\usepackage[hidelinks, bookmarks=false]{hyperref}
\usepackage{tabularx}
\usepackage{enumitem}
\usepackage{balance}

\usepackage{todonotes}
\usepackage{listings}
    
\usepackage{blindtext}
\usepackage{booktabs} 
\usepackage{dirtytalk} 

\begin{document}

\title{Integrating Static Code Analysis Toolchains\thanks{This work was funded by German Federal Ministry of Education and Research (BMBF) under grant \#01IS15031A as part of ASSUME~\cite{assume} project.}}

\author{
    \IEEEauthorblockN{
        \href{mailto:mkern@fzi.de}{Matthias Kern}\IEEEauthorrefmark{1},
        \href{mailto:ferhat.erata@yale.edu}{Ferhat Erata}\IEEEauthorrefmark{4}\IEEEauthorrefmark{5},
        \href{mailto:markus.iser@kit.edu}{Markus Iser}\IEEEauthorrefmark{3},
        \href{mailto:carsten.sinz@kit.edu}{Carsten Sinz}\IEEEauthorrefmark{2},
        \href{mailto:floiret@kth.se}{Frederic Loiret}\IEEEauthorrefmark{3},
        \href{mailto:otten@fzi.de}{Stefan Otten}\IEEEauthorrefmark{1}, and
        \href{mailto:sax@fzi.de}{Eric Sax}\IEEEauthorrefmark{1},
    }

\IEEEauthorblockA{
    \IEEEauthorrefmark{1}FZI Research Center for Information Technology, Karlsruhe, Germany}
\IEEEauthorblockA{
    \IEEEauthorrefmark{2}Karlsruhe Institute of Technology, Institut f\"{u}r Theoretische Informatik, Karlsruhe, Germany}
\IEEEauthorblockA{
    \IEEEauthorrefmark{3}KTH Royal Institute of Technology, Embedded Control Systems, Stockholm, Sweden}
\IEEEauthorblockA{
    \IEEEauthorrefmark{4}Yale University, Department of Computer Science, New Haven, USA}
\IEEEauthorblockA{
    \IEEEauthorrefmark{5}UNIT Information Technologies, Research \& Development, Izmir, Turkey}
\textit{\IEEEauthorrefmark{1}\{mkern, otten, sax\}@fzi.de \IEEEauthorrefmark{2}\{markus.iser, carsten.sinz\}@kit.edu \IEEEauthorrefmark{3}floiret@kth.se \IEEEauthorrefmark{4}ferhat.erata@yale.edu}
}    

\maketitle

\begin{abstract}
This paper proposes an approach for a tool-agnostic and heterogeneous static code analysis toolchain in combination with an exchange format. This approach enhances both traceability and comparability of analysis results. State of the art toolchains support features for either test execution and build automation or traceability between tests, requirements and design information. Our approach combines all those features and extends traceability to the source code level, incorporating static code analysis. As part of our approach we introduce the \say{ASSUME Static Code Analysis tool exchange format} that facilitates the comparability of different static code analysis results.
We demonstrate how this approach enhances the usability and efficiency of static code analysis in a development process. On the one hand, our approach enables the exchange of results and evaluations between static code analysis tools. On the other hand, it enables a complete traceability between requirements, designs, implementation, and the results of static code analysis. Within our approach we also propose an OSLC specification for static code analysis tools and an OSLC communication framework.
\end{abstract}

\begin{IEEEkeywords}
Traceability, Interoperability, Static Analysis, OSLC
\end{IEEEkeywords}
\section{Introduction} 
\label{sec:intro}
Highly automated vehicles with more than hundred electrical control units (ECUs) and millions of lines of code are good examples of safety-critical, complex systems \cite{100ECUs, LOC}. In the future, with the technological advances in autonomous driving, the complexity of those highly automated mobility systems will increase further in the number of sensors, communication pathways, and functionality. 

Yet, there are many tools without any linkage to other ones building so-called \say{islands of information}~\cite{Elk2016}. This means the data produced by such tools has no traceable connection between each other. In order to support the development of highly automated mobility systems in a safe and secure manner, toolchains that give the possibility to trace and exchange all design artifacts over the complete life-cycle are needed. Such toolchains would enable a direct exchange of information between different tools and thus enhance the traceability between the different sources of information. Standardized exchange formats are crucial to the creation of tool adapters which increase the interoperability among them and enhance the comparability of their outputs. Toolchains that are not constrained to a specific set of tools and that support the replacement of them, are so-called tool-agnostic toolchains.

Today, it is difficult to reuse configurations and to compare reports of different static code analysis tools (SCAT) since they mostly use a proprietary data format for both the analysis results and the analysis configuration. Furthermore, they have its own strengths and developers must often aggregate the analysis results of different analyzers to form an overall picture of program quality~\cite{ComparativeStudy}. However, without standardized exchange formats, it is not easy to combine their strengths. Additionally, it is very common for tool vendors to offer linkage between their own products, especially for web-based ones; nevertheless, tools of different vendors are required within a toolchain. Besides, there are many tools without a possibility to link their data easily between each other. To overcome aforementioned limitations, we propose a tool-agnostic and heterogeneous toolchain for SCAT.

The rest of the paper begins by presenting the background and related-work of our approach in Section~\ref{sec:related}. In Section~\ref{sec:overview} we give an overview of concepts through an exemplary use-case for \say{ASSUME Static Code Analysis Tool Exchange Format (ASEF)}, developed within the European ASSUME ITEA3 Project~\cite{assume}. In Section~\ref{sec:technical} we introduce technical concepts implementing the toolchain in which an adapter communication framework based on OSLC and the ASEF Format are presented, as well as an approach for static code analysis automation that supports traceabilty to design artifacts. Finally we give a conclusion and future work in Section~\ref{sec:conclusion}.

\section{Background and Related Work}
\label{sec:related}
In this section, we present the basic concepts and technologies that are necessary to understand our work. In Section~\ref{subsec:static}, we give a quick introduction into static code analysis with three small examples to motivate the necessity of static code analysis. In Section~\ref{subsec:traceability}, we describe the current literature on the traceability research to position our work in this area. In section~\ref{subsec:oslc}, we present OSLC, which builds technologically the base of our approach and is used to set up our toolchain that addresses system development. Since one of the main contributions of this work is the so-called ASSUME SCA tool exchange format (ASEF), in Section~\ref{subsec:sarif}, we explain a closely related standard, ``Static Analysis Results Interchange Format'' (SARIF) and discuss how our format is complementary to the SARIF. Finally, in Section~\ref{subsec:ci}, we present common continuous integration (CI) technique that today does not support traceability to design artifacts and comparability of different static code analysis tools. 

\definecolor{shadecolor}{RGB}{190,190,190}
\newcommand{\mybox}[1]{\par\noindent\colorbox{shadecolor}
{\parbox{\dimexpr\linewidth-6\fboxsep\relax}{#1}}}

\subsection{Static Analysis} \label{subsec:static}

Static code analysis involves methods and algorithms to automatically proof the absence of specific types of unwanted behavior in a piece of code. 
Static analysis tools can calculate a combination of input parameters and an execution trace that lead to an invalid state in a program.
Such invalid states might include \emph{undefined behavior}, the violation of \emph{custom assertions} or the access of \emph{uninitialized memory}.

\subsubsection{Undefined Behavior}
Unspecified semantics where the behavior of a programming language becomes unpredictable are commonly known as \emph{undefined behavior}. Such states are unwanted and should not be reachable at all. In the piece of code shown in example~\ref{algo:undef}, the removal of lines~2 and~3 leads to the reachability of an undefined state (considering the semantics of the C programming language).

\vspace*{-0.5em}
\begin{algorithm}
\SetAlgorithmName{Example}{example}{List of Examples}
\DontPrintSemicolon
$\mathsf{int~} z \leftarrow a - b$\;
\If {$z = \mathsf{MIN\_INT}$} { handle\_invalid\_state() \; }
$\mathsf{int~} y \leftarrow -z$\;
\caption{Undefined Behavior}\label{algo:undef}
\end{algorithm}
\vspace*{-0.5em}

\subsubsection{Custom Assertions}
Code optimization can lead to obfuscated pieces of code that are hard to comprehend and verify. 
There are scenarios, where code optimization is indispensable. 
Example~\ref{algo:equiv} shows how developers can use a \emph{custom assertion} to use static analysis to show that an optimized piece of code still behaves exactly the same as the unoptimized variant.

\vspace*{-0.5em}
\begin{algorithm}
\SetAlgorithmName{Example}{example}{List of Examples}
\DontPrintSemicolon
$\mathsf{int~} a \leftarrow \mathsf{foo}()$\;
$\mathsf{int~} b \leftarrow \mathsf{foo\_optimized}()$\;
$\mathsf{static\_assert}(a = b)$\;
\caption{Function Equivalence}\label{algo:equiv}
\end{algorithm}
\vspace*{-0.5em}

\subsubsection{Uninitialized Memory}
Working with old or unstructured low-level code can be a challenge with respect to memory management.
Functional extensions might lead to non-trivial bugs which can not easily be discovered.
Example~\ref{algo:uninit} is an abstract representation of a common situation.

\vspace*{-0.5em}
\begin{algorithm}
\SetAlgorithmName{Example}{example}{List of Examples}
\DontPrintSemicolon
\If {init-condition} {
    initialize memory\;
}
do some processing and access memory\;
\caption{Access Uninitialized Memory}\label{algo:uninit}
\end{algorithm}
\vspace*{-0.5em}

While \texttt{init-condition} (see line~1) might hold whenever needed in the original version of the software this property might get lost during a sequence of extensions and patches.
Static code analysis tools can automatically trace execution paths to states where uninitialized memory is accessed.

\subsection{Traceability} \label{subsec:traceability}

Regarding the industrial tools and technologies on traceability, modeling tools such as EMF \cite{steinberg2008emf} and SysML \cite{SysML}, requirement interchange standard (ReqIF \cite{ReqIF}) and management tools such as RMF\footnote{\url{https://www.eclipse.org/rmf/}} and IBM Rational DOORS \cite{doors2011requirements} provide some automated or manual means to specify and manage traceability. However, none of them provides integration with static analysis code analysis tools, especially on a heterogeneous development and design environment.

\subsection[OSLC]{Open services for Lifecycle Collaboration (OSLC)} \label{subsec:oslc}
Open services for Lifecycle Collaboration (OSLC)\footnote{\url{http://open-services.net/software/}} is an open community that defines specifications to link the data of different tools, used in the Application Lifecycle Management (ALM)~\cite{ALM2008} and Product Lifecycle Management (PLM)~\cite{PLM2009}, in order to directly support traceability. The OSLC specifications build on REST~\cite{REST}, the W3C Resource Description Framework (RDF) and Linked Data\footnote{\url{http://open-services.net/}}.

OSLC offers specifications for the requirement-management, the quality-management (QM) and the architecture-management. With the architecture-management specification, data from modeling-tools can be mapped to resources. Data from testing tools can be mapped with the help of the QM specification~\cite{OSLCSpec}. However, a specification for mapping results of static code analysis tools is missing. Therefore, a resource definition based on the QM specification and the ASEF format have been created as part of our approach. This specification is shown more in detail in the Section~\ref{subsec:specification}.

For several tools, there are already OSLC adapters available, like the Matlab Simulink integration from Axel Reichwein\footnote{\url{https://github.com/ld4mbse/oslc-adapter-simulink}} or for Bugzilla, a bug-tracking tool\footnote{\url{http://wiki.eclipse.org/Lyo/BuildOSLC4JBugzilla}}.

\subsection[SARIF]{Static Analyis Results Interchange Format (SARIF)} \label{subsec:sarif}
The Static Analysis Results Interchange Format (SARIF) is a standardized interchange format that enables the aggregation of results of different analysis tools. SARIF was originally developed by Microsoft and is currently being standardized by OASIS in the OASIS SARIF Technical Committee. The format addresses a variety of analysis tools that can indicate problems related to program qualities such as correctness, security, performance, compliance with contractual or legal requirements, compliance with stylistic standards, understandability, and maintainability. There are several tools available for the programming language C\# as SDKs or Converters. Furthermore, there is a Viewer for Visual Studio extension \footnote{\url{https://sarifweb.azurewebsites.net/}}. SARIF enhances the usability by combining and comparing the result in a more easier way than the several competing static analysis tools \footnote{\url{https://www.oasis-open.org/committees/tc_home.php?wg_abbrev=sarif}}.

\subsection[CI]{Continuous integration with static code analysis tools} \label{subsec:ci}
Current static code analysis tools like Astr\'ee\footnote{\url{https://www.absint.com/astree/index.htm}} from AbsInt or Coverty Scan from Synopsys offer a Jenkins\footnote{\url{https://jenkins.io/}} plugin that allows using these tools before the build process starts. Jenkins itself enables a continuous integration. The integration process, including testing and a build process, is normally triggered by a REST request. This request is normally sent out from a Git-repository manager, like GitHub or GitLab, after code was pushed to a specific git repository. However, traceability from testing results to requirements or other design artefacts is not supported through Jenkins. Furthermore, Jenkins plugins are specific for each tool, so there is no standard that enables a plugin for different static code analysis tools and offers comparability between different analysis tools.

\section{Concept and Use-Case} \label{sec:overview}

\begin{figure}[ht]
\centering
\includegraphics[width=\linewidth]{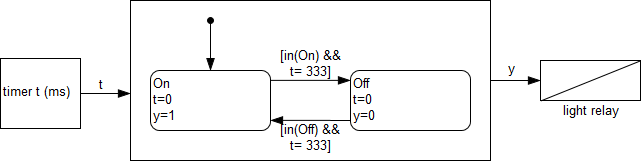}
\vspace*{-1.0em}
\caption{Direction indicator lamp for a car.}\label{fig:direction_lamp}
\end{figure}

\begin{figure}[ht]
\centering
\vspace*{-0.5em}
\includegraphics[width=\linewidth]{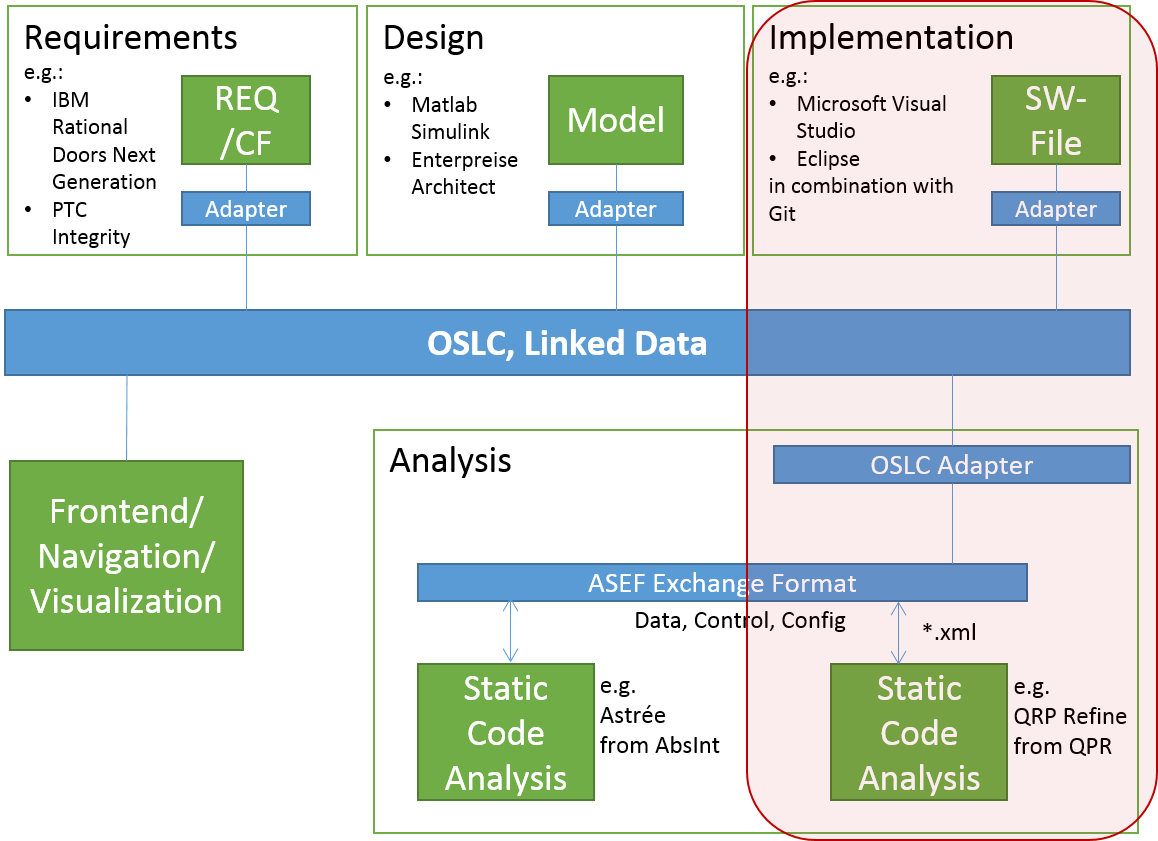}
\vspace*{-1.5em}
\caption{Toolchain}\label{fig:toolchain}
\vspace*{-0.5em}
\end{figure}

To motivate our approach we present an illustrative example for the development of a functionality for a direction indicator lamp.  This functionality could be run on an electronic control unit (ECU). Direction indicator lamps or informally ``blinkers'' or ``flashers'' are blinking lamps mounted near the left and right front of a car and can be activated by the driver at a time to advertise intent to turn or change lanes towards that side. For direction indicator lamps exist regulations like the \emph{E/ECE/324/Rev.1/Add.47/Rev.12 -  E/ECE/TRANS/505/Rev.1/Add.47/Rev.12} \footnote{\url{https://www.unece.org/fileadmin/DAM/trans/main/wp29/wp29regs/2015/R048r12e.pdf}}. To find all necessary regulations and requirements, requirement engineering tools as shown in Figure~\ref{fig:toolchain} can be used. In this regulation there is a requirement in section 6.5 that says \emph{\say{The light shall be a flashing light flashing 90$\pm$30 times per minute}}. To fulfill this requirement the functionality of the direction indicator lamp is designed with a \textit{stateflow chart} (cf. Figure~\ref{fig:direction_lamp}). The design of the functionality could be done with design tools as shown in Figure~\ref{fig:toolchain}.

 After the system design the code can be written or generated during the implementation activity. A piece of software code for the functionality of the direction indicator lamp could look like that shown in Listing~\ref{algo:indicatorLamp}.
 
 Here, we assume to have a hardware timer that is increased every millisecond by one,
and its value can be read out using function \texttt{getTimer()}, which returns a
\textsf{signed short}. To achieve a blinking frequency of 1.5 Hz, the light should be
switched every 333 ms. To make sure that the implementation satisfies the requirement,
we have added assertions (via \texttt{static\_assert}) to make sure that a switch
occurs after between 250 ms and 500 ms and that the time increases in each iteration of the loop.

\lstset{
    language=C, 
    commentstyle=\color{green}, 
    keywordstyle=\color{blue}, 
    basicstyle = \ttfamily \color{black} \footnotesize, 
    numbers=left, 
    numberstyle=\tiny,
    caption = {Direction indicator lamp C Code.},
    captionpos = b, 
    label = algo:indicatorLamp,
    xleftmargin=1.0em,
} 

\begin{lstlisting} 
typedef enum { Off, On } state_t;
typedef short time_t;

extern time_t getTimer();
extern void setIndicatorLamp(state_t s);

int timerExpired(time_t start, time_t end, time_t diff)
{
    return end-start > diff;
}

void process()
{
    time_t startTime = getTimer(), currentTime;
    state_t light = Off;
    while (1) {
        currentTime = getTimer();
        static_assert(currentTime - startTime >= 0);
        if (!timerExpired(startTime, currentTime, 333)){
            continue;
        }
        if (light == Off) {
            setIndicatorLamp(light = On);
        } else {
            setIndicatorLamp(light = Off);
        }
        static_assert(currentTime - startTime >= 250);
        static_assert(currentTime - startTime <= 500);
        startTime = currentTime;
    }
}
\end{lstlisting}

Due to an integer overflow bug, the implementation will not work in all cases, in particular
on a 32-bit platform.
E.g., if \texttt{startTime = 32700} and \texttt{currentTime} has a negative value
due to an overflow, then \textsf{end - start} in function \texttt{timerExpired()} will be 
a large negative value (due to integer promotion no overflow occurs in the subtraction),
and it will take a long time (approx. 65 seconds) until a timer expiration is reported.

This kind of overflow bug is not only of academic interest, but also of practical importance,
as an incident from 2015 shows: Engines of the Boeing 787 Dreamliner could fail due to loss of electric power after 248 days of continues
operation\footnote{See, e.g., \url{https://arstechnica.com/information-technology/2015/05/boeing-787-dreamliners-contain-a-potentially-catastrophic-software-bug}.}. The fault was caused by a timer-related integer overflow
bug similar to that of the example above.

The analysis of the code from Listing~\ref{algo:indicatorLamp} belongs to analysis activities, and many static analysis tools will be able to find the bug in the implementation. Output from an analysis tool might look as in Example~\ref{algo:analysisResult}.
%
\begin{algorithm}
\SetAlgorithmName{Example}{example}{List of Examples}
\DontPrintSemicolon
Assertion in line 18 failed: \\
    {startTime = 32452} \\
    {currentTime = -32684}
\caption{Analysis results.}\label{algo:analysisResult}
\end{algorithm}

All these artifacts of our example should be traceable within the proposed toolchain. Therefore we give here an exemplary overview of possible tools which allows us to manage and produce the necessary results. An overview of this toolchain is given in Figure~\ref{fig:toolchain}. To make a proof of concept we implemented a part of this concept (red shaded box in Figure~\ref{fig:toolchain}). As a technology to implement the traceability links among artifacts we employ Open Services for Lifecycle Collaboration (OSLC) (cf. Section~\ref{sec:related}). 
The requirement analysis activities can be supported with IBM Rational Doors Next\footnote{\url{https://www.ibm.com/us-en/marketplace/rational-doors}} or PTC Integrity\footnote{\url{https://www.ptc.com/en/products/plm/plm-products/}}. The design activity could be supported with Matlab Simulink from MathWorks\footnote{\url{https://www.mathworks.com/products/simulink.html}} or Enterprise Architect from SparxSystems\footnote{\url{https://www.sparxsystems.eu/start/home/}}. With Matlab Simulink, the behaviour of systems can be modeled and simulated. The architectural description of a system can be described with Enterprise Architect, which supports Unified Modeling Language (UML) and Systems Modeling Language (SysML). For the software implementation we suppose an integrated development environment like Visual Studio from Microsoft or Eclipse from the Eclipse Foundation. For the static code analysis, Astr\'ee from AbsInt or QPR Refine from QPR Technologies could be used. Both tools use abstract interpretations of C code to detect runtime errors, data races or assertion violations and includes a MISRA C checker\footnote{\url{https://www.misra.org.uk/Activities/MISRAC/tabid/160/Default.aspx}}. In the box \say{Analysis} in Figure~\ref{fig:toolchain} two static code analysis tools can use the ASEF format (cf. Section~\ref{sec:asef}) to produce comparable results. However, only one analysis tool is adapted here in our toolchain.

\section{Technical Concepts}
\label{sec:technical}
To show the process behind the scenes of our toolchain the most important technical concepts are presented in this section. These concepts include the framework of our toolchain, the ASEF exchange format and a specification for static code analysis tools based on the OSLC quality management specification.

\subsection[framework]{The OSLC adapter communication framework}
\begin{figure}[ht]
\centering
\vspace*{-1.0em}
\includegraphics[width=\linewidth]{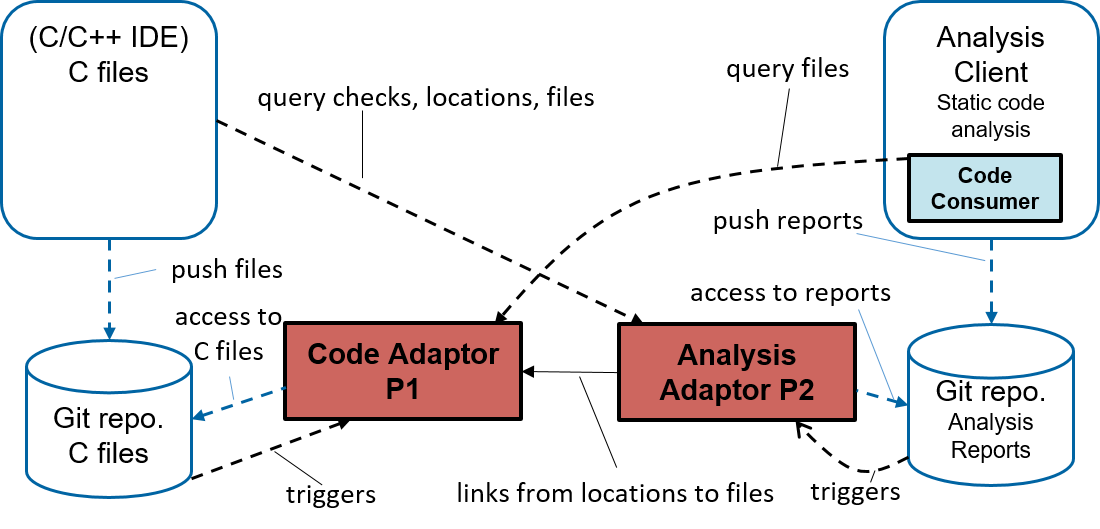}
\vspace*{-1.5em}
\caption{Communication framework}\label{fig:framework}
\vspace*{-1.3em}
\end{figure}

In the following section, the framework and its communication is described. Here, only the communication between an integrated development environment (IDE) and a static code analysis tool is regarded (see red box in Figure \ref{fig:toolchain}).
In Figure \ref{fig:framework}, the communication between the IDE and the static code analysis tool is depicted. The static code analysis tool uses a client that manages the communication.  For each communication participant exists a git repository. Within these git repositories, the data is stored and version managed in a standardized format. In the case of C code the standardized format is the C code itself. In the case of the static code analysis, the analysis report is stored in the tool independent ASEF exchange format. Through the git repository, only checked in versions are linked within the tools in the toolchain. The adapters \say{Code Adapter P1} and \say{Analysis Adapter P2} parses the data from the git repositories into the Linked Data format namely \say{Resource Description Framework} (RDF). The adapters are triggered every time a new version is pushed into the corresponding git repositories. During parsing, they link the corresponding information of the different adapters. In this case, the information of the analysis report is linked with the C code. The participants can retrieve the linked information via the both adapters. The adapters were implemented with the help of the model based development tool \say{Lyo Modeller}.

\subsection[ASEF]{The ASEF Format}
\label{sec:asef}
The ASSUME SCA tool exchange format (ASEF) offers an XML schema for a tool-agnostic configuration format in static code analysis and for the reporting of analysis results. 
We developed ASEF Format aiming at tool-interoperability and facilitating well-defined check-semantics. 
The format is extensible and allows tool-dependent configuration. The schemes and the documentation are available on the following pages:
\begin{center}
\small{\fbox{\bf{\url{http://assume-project.github.io/download.html}}}}
\end{center}

\subsubsection{ASEF Configuration Format}
\label{sec:asef-config}
The ASEF configuration is split into a global part and a local part, in which the global part is intended to contain the main configuration that can be shared across multiple hosts, 
whereas specific hosts are able to adapt the local part of the configuration to their needs. 
Listing~\ref{lst:asef} shows a small example of an ASEF configuration. 

The configuration allows the definition of \emph{source modules}, \emph{hardware targets}, \emph{language targets} and \emph{check targets}. 
Source modules define the files to analyze. Hardware targets define hardware specific properties such as pointer size or endianness. Language targets define details about the language standard and system to be checked. 

\lstset{language=xml, tabsize=1, frame=lines,
caption=ASEF Example Configuration, label=lst:asef,
keywordstyle=\color{blue}\bf,commentstyle=\color{gray},stringstyle=\color{OliveGreen}\it,
breaklines=true,basicstyle=\scriptsize, numbers=none, xleftmargin=-0.5em,
emph={HardwareTarget, LanguageTarget, CheckTarget, CorrectnessCheckCategory, ExecutionModelTarget, SourceModule, SourceFile, AnalysisTask, URISubstitutionRule},
emphstyle={\color{Bittersweet}\bf}}
\begin{lstlisting}
<asef:Configuration xsi:schemaLocation="ASC3F.xsd">
<asef:GlobalConfiguration>
<asef:CommonConfiguration>
  <HardwareTargets>
   <HardwareTarget xsi:type="asef:HomogenousHardwareTarget" name="generic32" endianness="big" pointerSize="32"/>
  </HardwareTargets>
  <LanguageTargets>
    <LanguageTarget xsi:type="asef:CLanguageTarget" name="basicC11" standardRevision="C11" />
    </LanguageTarget>
  </LanguageTargets>
  <CheckTargets>
    <CheckTarget xsi:type="asef:CCheckTarge" name="base">
      <CorrectnessCheckCategory name="assert"/>
      <CorrectnessCheckCategory name="numeric"/>
      <CorrectnessCheckCategory name="controlflow"/>
      <CorrectnessCheckCategory name="mem"/>
    </CheckTarget>
  </CheckTargets>
  <ExecutionModelTargets>
    <ExecutionModelTarget xsi:type="asef:CSynchronousExecutionModelTarget" name="sync">
      <EntryPoints>
        <EntryPoint>main</EntryPoint>
      </EntryPoints>
    </ExecutionModelTarget>
  </ExecutionModelTargets>
  <SourceModules>
    <SourceModule name="main" rootUri="$Repository$">
    <SourceFiles>
      <SourceFile uri="example.c" id="1"/>
    </SourceFiles>
    </SourceModule>
  </SourceModules>
  <AnalysisTasks>
    <AnalysisTask name="analyzeMainSourceModule">
      <HardwareTarget>generic32</HardwareTarget>
      <SourceModule>main</SourceModule>
      <LanguageTarget>basicC11</LanguageTarget>
      <CheckTarget>base</CheckTarget>
      <ExecutionModelTarget>sync</ExecutionModelTarget>
    </AnalysisTask>
  </AnalysisTasks>
</asef:CommonConfiguration>
</asef:GlobalConfiguration>
<asef:LocalConfiguration>
  <URISubstitutionRules>
    <URISubstitutionRule token="$Repository$" substitution="/local/path/to/repositories">
</URISubstitutionRules>
</asef:LocalConfiguration>
</asef:Configuration>
\end{lstlisting}

Via check targets, several check categories define which checks should be executed. 
One of the key-strength of ASEF lies in the precise specification of the check semantics. 
As different static analysis tools offer different stages of precision in various check categories there was a need to define these levels of precision. 
They are used in the ASEF Reports when a flaw was discovered. 
More details about check semantics can be found in Section~\ref{sec:asef-report}.

The execution of checks is triggered through the definition of \emph{analysis tasks}.
Analysis tasks are a combination of the previously defined targets, they specify which checks should be executed with which hardware and language configuration etc.

\subsubsection{ASEF Report Format}
\label{sec:asef-report}
The analysis reports involve source locations, failure traces, and check semantics.

Locations define the row and column where a fault was detected and are assigned to the checks via identifiers. It is also possible to refer from a location to another. 
This is useful to describe so-called macro locations, because macros can use code from other files or locations.
 
A check status can be safe, unsafe, undecided, warning, and syntactic violation. 
The check category describes the kind of fault. Table~\ref{tab:check-semantics} shows a small excerpt of the hierarchy of ASEF check categories and how the categories map to those of various static analysis tools. 
ASEF offers well-defined check semantics and thus enables comparability of the results of the different static analysis tools. 

\begin{table}[ht]
\vspace*{-1.0em}
\caption{ASEF Check Semantics vs. Native Categories of Various Tools (including Polyspace (PS) check categories, QPR check categories and Astree (AS) alarm categories)}
\label{tab:check-semantics}
\vspace*{-0.5em}
\centering
\begin{tabular}{@{}l p{5.3cm}@{}}
\toprule
{ASEF Category}  & {Native Categories}\\
\midrule
numeric.overflow        & {PS:OVFL}, {AS:"Overflow in arithmetic"}, {AS:"Initializer range"} \\
numeric.overflow.int    & {QPR:arithmetic.overflow}, {QPR:shift.overflow}\\
numeric.shift           & {PS: SHF} \\
numeric.shift.rhs       & {AS:"Wrong range of second shift argument"}\\
numeric.shift.rhs.amount & {QPR:shift.by.amount}\\
numeric.shift.rhs.negative & {QPR:shift.by.negative}\\
\midrule
mem                         & {PS:COR} \\
mem.ptr.deref               & {PS:IDP}, {QPR:pointer.dereference} \\
mem.ptr.deref.misaligned    & {AS:"Dereference of mis-aligned pointer"}\\
mem.ptr.deref.invalid       & {AS:"Dereference of null or invalid pointer"}\\
mem.ptr.deref.field         & {AS:"Incorrect field dereference"} \\
\bottomrule
\end{tabular}
\end{table}

\subsection[specification]{Proposal of an OSLC specification for static code analysis data and results exchange} \label{subsec:specification}
\begin{figure}[ht]
\centering
\includegraphics[width=\linewidth]{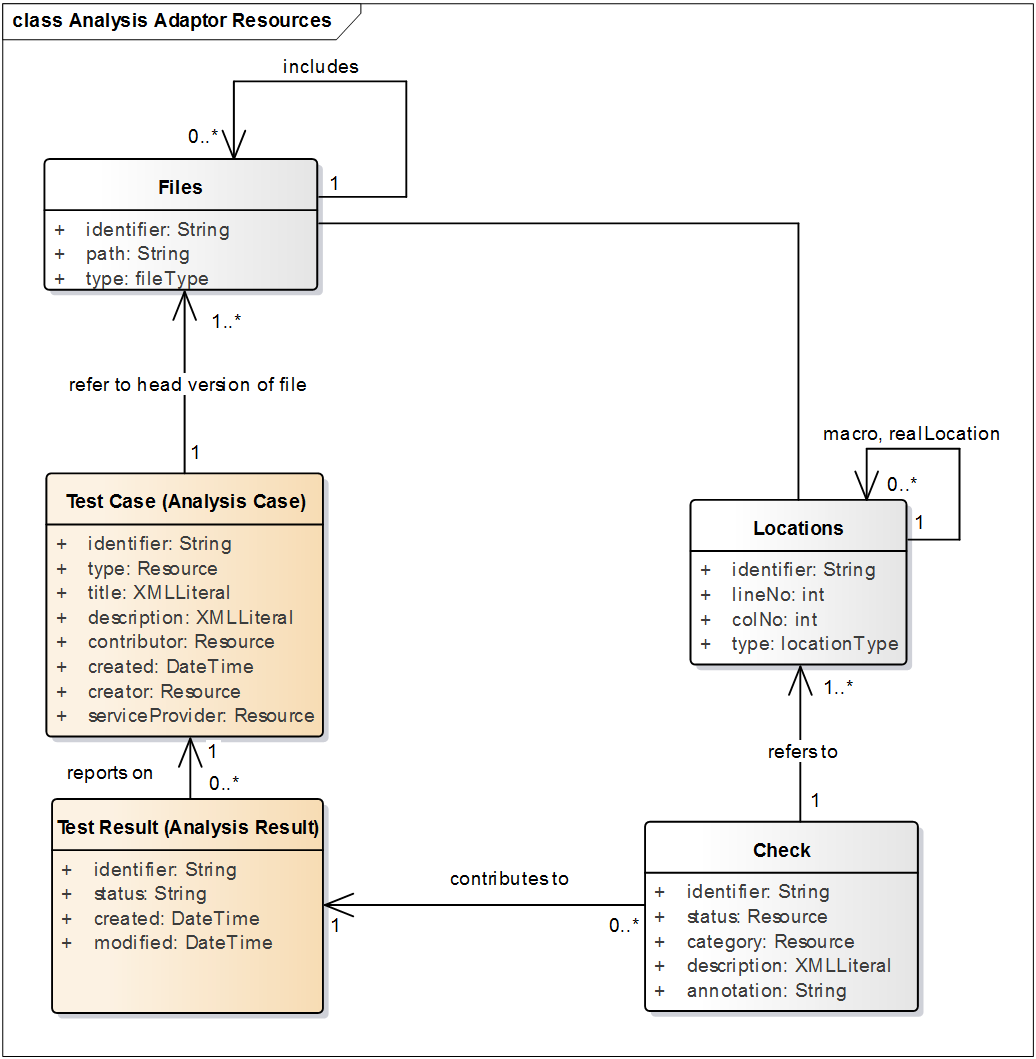}
\vspace*{-1.3em}
\caption{Proposal static code analysis OSLC-Specification}\label{fig:specification}
\vspace*{-1.5em}
\end{figure}

Figure~\ref{fig:specification} shows the resource definition based on both the OSLC quality management (QM) specification and the ASEF format. This definition was used to create the OSLC adapter to link the analysis information with the files and to offer an analysis case that enables the configuration for the static code analysis via OSLC. In the following, the proposed specification is explained in detail.

The orange boxes in Figure~\ref{fig:specification} represent resources that are adapted from the OSLC QM specification. The new name of the adapted resources stands in brackets behind the original name. The grey boxes represent resources based on the ASEF format. In the following, these resources are explained more in detail.

\texttt{The Analysis Case} is linked with all files that should be analyzed and contains the configuration for the static code analysis. \texttt{The Analysis Result} bunches all so-called Checks of one ASEF analysis report. There can be different Analysis Results for different versions of the source code files. One \texttt{Analysis Result} refers to a specific \texttt{Analysis Case}. \texttt{Checks} contain the type of a detected fault at a specific location in the code. There is a link to the location, where the fault occurs. A \texttt{Location} gives the line and column number in the code where a problem or fault was located. The \texttt{Location} is linked with the appropriate file, or with another \texttt{Location} in the case that the \texttt{Location} is a so-called macro. Through the presented linkage the \texttt{Analysis Cases} are traceable from their results up to the source code of a specific version.

\subsection[automation]{Approach for a static code analysis automation process}
In this section, an approach for an automation process of the static code analysis within the toolchain is proposed. This process is oriented at the communication framework (cf. Figure ~\ref{fig:framework}) and is divided into the ten following steps:
1)	A developer push files from an IDE into a git repository;
2)	GitLab, that manages this git repository, triggers the analysis Client via a so-called webhook;
3)	The Analysis client requests the new files from the Code Adapter P1, stores them locally and looks which analysis cases are affected through the changed files of the current commit;
4)	The Analysis Client runs for each affected \texttt{Analysis Case} the analysis;
5)	The Analysis Client reads the configuration from the \texttt{Analysis Cases} and changes the path for the analysis to the local stored C language files;
6)	The Analysis Client starts the static analysis tool via an API;
7)	The static code analysis tool analysis the files defined in the analysis configuration and produces an analysis report in the ASEF format and sends a ready acknowledgment to the analysis client;
8)	The Analysis Client replaces in the analysis report the local files with the Unified Resource Identifier (URI) to the files in the Code Adapter P1 and pushes the report into the analysis repository;
9)	GitLab triggers the analysis adapter, after the analysis client pushed the analysis report into the analysis repository;
10) The analysis adapter parses the information of the analysis report into Linked Data resources and links the resources of the \texttt{Locations} with the \texttt{Files} provided by the Code Adapter P1. Afterwards the developer can use a front-end to get a quick overview of the analysis results (cf. Figure ~\ref{fig:toolchain})

\section{Conclusion and future work}
\label{sec:conclusion}
In this paper we introduced the necessity for improved traceability from the requirements downwards to static code analysis within a heterogeneous and tool-agnostic toolchain. Starting with the integration of our approach into the V-model, the state of the art and an overview about related work, we show the gaps of common solutions and motivate our approach with an use-case of a \textit{direction indicator lamp}. 

The approaches to enhance the traceability and the usability are presented as well as a technical concept with a new static code analysis exchange format, namely ASEF, a communication framework, and an OSLC specification to implement an adapter for static code analysis tools. Up to now, our implementation create linkages between code and static analysis results. For future work we are going to add new tools to prove the usability and traceability of our concepts. We demonstrated that the ASEF format brings several advantages such as the possibility to configure static code analysis tools in a uniform way. Nevertheless, we aim to show whether our approach can work also with the SARIF Standard to reach a larger community.

\bibliographystyle{IEEEtran}
\bibliography{literature}

\end{document}